\documentclass[12pt,a4paper]{article}

\usepackage{amsmath}
\usepackage{amsfonts}
\usepackage{amsthm}

\def\C{\mathbb C}
\def\R{\mathbb R}
\def\dom{\mathrm {dom}\,}
\def\LL{\mathrm L}
\def\CC{\mathrm C}
\def\Id{\mathbf 1}

\newcommand{\dis}{\displaystyle}
\newcommand{\var}{\varepsilon}
\newcommand{\la}{\langle }
\newcommand{\ra}{\rangle }

\newcommand{\hil}{\mathcal H}
\newcommand{\wseta}{\rightharpoonup}
\newcommand{\wgconv}{\stackrel{\mathrm{W\Gamma}}{\longrightarrow}}
\newcommand{\sgconv}{\stackrel{\mathrm{S\Gamma}}{\longrightarrow}}

\newtheorem{teor2}{Theorem}

\newtheorem{prop2}{Proposition}

\newtheorem{obs2}{Remark}

\theoremstyle{definition}

\begin{document}

 \title{Mathematical predominance of Dirichlet  condition for the one-dimensional Coulomb potential}
\author{C\'esar R. de Oliveira {\small and} Alessandra A. Verri\\ 
\vspace{-0.6cm}
\small
\em Departamento de Matem\'{a}tica -- UFSCar, \small \it S\~{a}o Carlos, SP,
13560-970
Brazil\\ \\}
\date{\today}

\maketitle 

\begin{abstract}  We restrict a quantum particle under a coulombian potential  (i.e., the Schr\"odinger operator with  inverse of the distance potential) to three dimensional tubes along the $x$-axis and diameter $\varepsilon$, and study the confining limit $\varepsilon\to0$. In the repulsive case we prove a strong resolvent convergence to a one-dimensional  limit operator, which presents Dirichlet boundary condition at the origin. Due to the possibility of the falling of the particle in the center of force, in the attractive case we need to regularize the potential and also prove a norm resolvent convergence  to the  Dirichlet operator at the origin.  Thus, it is argued that, among the infinitely many self-adjoint realizations of the corresponding problem in one dimension, the Dirichlet boundary condition at the origin is the reasonable one-dimensional limit.   
\end{abstract}

\section{Introduction} 

In this work we combine three questions that have been investigated in the last years: the reduction of dimension in physical models, the selection of a self-adjoint extension of a quantum hamiltonian, and the one-dimensional hydrogen atom. First, suppose one has a  quantum particle in a (three-dimensional) waveguide and under a singular potential so that, in three dimensions ({\sc3D}) the potential does not imply in more than one self-adjoint extension of the hamiltonian operator, but the corresponding problem in one dimension ({\sc1D}) has infinitely many self-adjoint extensions; so, the natural question is what of such extensions is selected under the process of confinement as the cross section diameter of the waveguide vanishes. Second, the problem we have chosen to study the previous question is the Schr\"odinger operator with Coulomb potential 
\begin{equation}\label{potVC}
V_C(\vec x)= -\frac{\kappa}{|\vec x|},\quad 0\ne \kappa\in\R,\;\vec x\in\R^n,
\end{equation} a problem that has a controversial, albeit absorbing, history in {\sc1D}. We will follow the nomenclature of some works and, independently of the dimension considered, we also call this model a {\em hydrogen atom}.  

Although our main results  on this singular potential are based on a careful mathematical analysis of this  problem,  we think a rather detailed discussion on its  physical and historical  aspects will consist of an additional motivation.

It is worth mentioning that the {\sc1D} hydrogen atom has been used:\begin{itemize} \item as a simplified  model  in theoretical and numerical studies \cite{JSS,DKS,LCO3}, particularly describing atoms in very intense magnetic fields \cite{RWHR}; 
\item in the description of electrons hovering above superfluids \cite{Care,CC,Cole};
\item  in investigations of the above threshold ionization of atoms under very intense laser fields \cite{PJiang}; 
\item  in applications to condensed matter physics \cite{Nieto,Care};
\item in descriptions of  electrons trapped in one-dimensional hydrogenic levels, which have been suggested as a possible device for quantum computing \cite{DPS}; 
\item it was noted that the electronic distribution  of atoms, in excited states  under time-periodically electric fields, can be modelled by the one-dimensional hydrogen atom \cite{IS,GBM}. 
\end{itemize} Furthermore,  some experimental evidences for the {\sc1D} hydrogen atom (see \cite{CC,Wong} and papers that cite these works) have also been reported, and,  recently, in \cite{RRYB} the authors present a discussion against the assertion that  the 1D Coulomb potential ``does not properly exist.''  All in all we conclude that  the one-dimensional hydrogen atom is not just  a question of purely academic interest, which is an additional stimulus for the search of  the physically reasonable boundary condition at the origin (if any) among the admissible ones.

 In 1959 Loudon \cite{Lou} published a work that  ``popularized'' the   {\sc1D} hydrogen atom, although this problem has been previously considered in 1928  by Vrkljan \cite{VRs,VRs2}. From the mathematical viewpoint, the initial operator is 
\begin{equation}\label{defHponto}
\dot H=-\frac{\hbar^2}{2m}\frac{d^2}{d x^2} + V_C(x), \quad \dom \dot H = \CC_0^\infty(\R\setminus\{0\}),
\end{equation}and rather recently it was realized that the main source of controversies is that $\dot H$ is not essentially self-adjoint \cite{FLM,deOV}. Relativistic versions of the {\sc 1D} hydrogen atom have also been considered \cite{VRs2,SL,Barton,Hall}.

Differently from the {\sc3D} version, for which one may start with the domain $\CC^\infty_0(\R^3)$ and get an essentially self-adjoint operator, in {\sc 1D}  one is forced to start with~\eqref{defHponto} and the Coulomb singularity needs the assignment of boundary conditions at the origin in order to get a self-adjoint realization.  $\dot H$ is hermitian but not
self-adjoint, and it turns out that it has deficiency index
$n_+=2=n_-$ and so an  infinite family of self-adjoint extensions, which are the
candidates for the (quantum) energy operator of the {\sc1D} hydrogen atom.  In Appendix~\ref{1dHatom} there is a description of all self-adjoint extensions of the one-dimensional hamiltonian \eqref{defHponto}.

On all occasions that an initial energy operator has more than one self-adjoint extension, it is natural to ask which one (if any) is predominant in some sense. And due to the highly arguable {\sc1D} hydrogen atom history, the question becomes even more compelling in this case. An intuitive first guess is to impose Dirichlet boundary condition at the origin, in case one would expect that the {\sc1D} Coulomb singularity should act as a barrier that does not allow the electron to pass through itself, but this point is not so straightforward, and in~\cite{deOV} the question of permeability of the origin versus the self-adjoint realization is also discussed. For instance, there are boundary conditions different from Dirichlet so that the origin becomes impermeable as well.

An argument for the adoption of the Dirichlet boundary condition at the origin  (see Appendix~\ref{1dHatom}) is that in the limit of the {\sc 1D} Schr\"odinger operator  (so that the Coulomb singularity has been regularized)
\[
-\frac{\hbar^2}{2m}\frac{d^2}{d x^2} -\frac{\kappa}{|x|+a},\qquad a>0 \; (\kappa>0),
\]as $a\to0$,  the Dirichlet boundary condition shows up in the norm resolvent limit $H_D$~\cite{Klaus}.  Other mathematical treatments related to the one-dimensional Coulomb potential can be found in \cite{Kurasov, NZ, Nenciu}.  

Here we present another mathematical argument in favor of Dirichlet, one we think is also more appealing from the physical viewpoint: since in practice a one-dimensional system is an approximation (under suitable conditions) of  three-dimensional ones,  we begin with the well-defined {\sc3D} Coulomb system restricted to a tube and take the (singular) confining limit to a line, and  next we give mathematical evidences supporting the prevalence of Dirichlet boundary condition. Our technical arguments and proofs will distinguish the  repulsive case (i.e., $\kappa<0$, with strong resolvent operator convergence) from the attractive  one ($\kappa>0$, with norm resolvent convergence of a sequence of regularized potentials~\eqref{regPot}).  From the physical viewpoint, in the attractive case there is the possibility of the particle to be captured by the center of coulombian force, and the  regularization of the potential is to control such possibility (represented by a divergence of energy to~$-\infty$). We will assume that the coulombian center of force is kept fixed at the origin during the confining process; a crucial step will be the introduction of suitably selected intermediary forms and operators.

Based sole on energy expectations, in~\cite{deO_PLA} one of the authors has  inferred that Dirichlet is the physical boundary condition at the origin for the one-dimensional hydrogen atom. Hence, this work should be considered a more complete argument towards the same physical conclusions, since an operator analysis is performed and  resolvent convergences to~$H_D$ are proven.

Recently, some  closely related results have appeared in~\cite{DST}; there it is shown that the two-dimensional Coulomb quantum Hamiltonian is obtained as a  resolvent limit of the Hamiltonian of a hydrogen atom in a planar slab as the width of the slab tends to zero. For a rather recent review and further references to the problem of Dirichlet waveguides, see \cite{krejc,LTW}.

 As mentioned above, by using a result from~\cite{Klaus} (restated in Theorem~\ref{ppp2} ahead), we obtain a convergence  with regularized potentials~\eqref{regPot} in the
attractive case, but such technical arguments  don't extend to the
repulsive case, for which we apply the technique of $\Gamma$-convergence~\cite{DalMaso,GammaConvT}. From now on we assume that $\hbar^2/(2m)=1$. 

In Section~\ref{sectSetup} we discuss details of the model we study as well as our main results (Theorems~\ref{teorAttractive} and~\ref{teorRepulsive}). The corresponding proofs appear in Sections~\ref{sectAttractive} and~\ref{sectRepulsive}.

\section{Setup and main results}\label{sectSetup} 
\subsection*{Hamiltonians}Let $\emptyset\ne S$ be an open and bounded subset  of $\R^2$,  homeomorphic to a disk, which will be transversely transported along the $x$-axis; as usual, denote by $\mathbf{i,j}$ and $\mathbf k$ the unit vectors pointing towards the positive directions of the axes $x,y_1=y$ and $y_2=z$, respectively. Assume that the point with coordinates $y_1=0=y_2$ belongs to~$S$ (this is convenient due to the Coulomb potential we will consider ahead), and that for any point $x\in\R$ the vector $\mathbf i$ is normal to~$S$. 

For $\varepsilon>0$, introduce the region
\[
\Omega_\alpha^\varepsilon := \left\{(x,y,z)\in\R^3: (x,y,z)=x{\mathbf i}+\varepsilon y_1 {\mathbf j}_\alpha(x) + \varepsilon y_2
{\mathbf k}_\alpha(x), (y_1,y_2)\in S \right \}
\]where, given  a $\CC^1$ function $\alpha:\R\to\R$ with $\alpha(0)=0$ and bounded derivative $\|\alpha'\|_\infty<\infty$ (the prime symbol $'$ will always indicate derivative with respect to the variable~$x$), 
\begin{eqnarray*} {\mathbf j}_\alpha(x) &=& \cos\alpha(x) {\mathbf j} -\sin \alpha(x) {\mathbf k}
\\ {\mathbf k}_\alpha(x) &=& \sin\alpha(x) {\mathbf j} +\cos\alpha(x) {\mathbf k}.
\end{eqnarray*}  The tube is then defined by the map $f_\alpha^\varepsilon:\R\times S\to
\Omega_\alpha^\varepsilon$, 
\begin{equation}\label{fvaralpha}
 f_\alpha^\varepsilon(x,y_1,y_2)=x{\mathbf i}+\varepsilon y_1 {\mathbf j}_\alpha(x) + \varepsilon y_2
{\mathbf k}_\alpha(x),
\end{equation}
and it represents a cylinder along the $x$-axis whose transverse section $S$, at position $x$, rotates by an angle $\alpha(x)$ with respect to the origin $x=0$. We define, for each $\phi \in {\cal H}_0^1(\Omega_\var)$, $\psi(x, y_1, y_2)= \phi(f_\alpha^\var(x, y_1, y_2))$.
From now on we will drop the index $\alpha$ and write simply $\Omega^\varepsilon$ for $\Omega_\alpha^\varepsilon$. The laplacian $-\Delta_{\Omega^\varepsilon}$ is taken with the Dirichlet boundary condition at the tube border; more precisely, it is the self-adjoint operator
acting in
 $\LL^2(\Omega^\varepsilon)$ associated with the positive sesqui\-lin\-e\-ar form, with domain $\hil_0^1(\Omega^\var)$,
\[ (\phi,\varphi)\longmapsto\la \nabla \phi,\nabla  \varphi\ra=\int_{\Omega^\varepsilon}\overline{\nabla  \phi}\,\nabla  \varphi\,dxdy\quad \phi,\varphi\in
\hil_0^1(\Omega^\var);
\]the inner product is in the space $\LL^2(\Omega^\varepsilon)$, $\nabla $ is the usual gradient in
cartesian coordinates $(x,y)=(x,y_1,y_2)$ and $dy=dy_1dy_2$. 

The operator we are interested in studying the confinement is, for each $\varepsilon>0$, the Schr\"odinger operator with Coulomb potential restricted to the tube $\Omega^\varepsilon\subset\R^3$ on putting Dirichlet boundary condition at the tube border; we begin with the self-adjoint operator $h^\varepsilon_\kappa$ whose quad\-rat\-ic form is
\[
\hil_0^1(\Omega^\varepsilon)\ni\phi\longmapsto \tilde b^\varepsilon_\kappa(\phi)= \|\nabla \phi\|^2 -\kappa \left\la \phi,\frac1{|(x,y)|}\phi\right\ra,
\]and  its action is given by
\[
h^\varepsilon_\kappa\phi = -\Delta_{\Omega^\varepsilon}\phi -\frac\kappa{|(x,y)|}\phi,\quad \phi\in \dom h^\varepsilon_\kappa = \hil_0^1(\Omega^\varepsilon)\cap  \hil^2(\Omega^\varepsilon).
\]

Under the  singular limit $\varepsilon\to0$, that is, when the tube is squeezed to the $x$-axis, by neglecting for a moment the eventual role played by the rotation of the cross section $\alpha(x)$, one formally expects to get a version of the  one-dimensional hydrogen atom
\[
h^\varepsilon_\kappa\phi\; \stackrel{\varepsilon\to0}{\longrightarrow}\; -\phi''(x) -\frac{\kappa}{|x|}\phi(x),
\] and our task is to make sense, under suitable conditions, of this limit, so defining a one-dimensional quantum energy operator that arises out of such confinement. Furthermore, we will show that the Dirichlet boundary condition at the origin will naturally be selected. However, we will need to regularize the potential if~$\kappa>0$, that is, instead of $-\kappa/|(x,y)|$ we will consider 
\begin{equation}\label{regPot}
-\frac{\kappa}{|(x,y)|+\var^\delta},
\end{equation} with $0<\delta<1/2$. We will write the corresponding quad\-rat\-ic forms in a suitable way.

As usual in the study of similar problems, we make an appropriate change of variables in order to simplify the region we work with, but the price we pay is a more complicated action of the subsequent quad\-rat\-ic forms and operators. 

Let $\lambda_0>0$ denote the first (which we also suppose to be simple) eigenvalue of the laplacian on~$S$, \[
-\Delta_\perp u_0=\lambda_0 u_0,
\] with positive eigenfunction $u_0\in \hil_0^1(S)$ and normalization $\int_S u_0(y)^2 dy=1$.  Here, the symbols $-\Delta_\perp, \nabla_\perp$ refer to the laplacian and gradient in the cross-section variables $y=(y_1,y_2)$, respectively; thus $\nabla  = (\partial_x,\nabla_\perp)$.   Since we are interested in the limit 
$\varepsilon\to0$, we perform two kinds of ``renormalizations;'' these are common approaches to balance singular problems,  and so to put them in a tractable form~\cite{DalMaso,BMT}.

When the tube is squeezed, there are divergent energies due (at least) to terms of the form
$\lambda_0/\varepsilon^2$ related to transverse
oscillations in the tube, and to the uncertainty principle with respect to the variables we want to eliminate. Hence we subtract
\[
\frac{\lambda_0}{\varepsilon^2}\|\phi\|^2
\]  from the quad\-rat\-ic form of $h^\varepsilon_\kappa$, and this is our first renormalization.
Now we perform the change of variables induced by the function $f_\alpha^\varepsilon$, so that the quad\-rat\-ic form $ \tilde b^\varepsilon_\kappa(\phi)-\frac{\lambda_0}{\varepsilon^2}\|\phi\|^2$ is written as
\begin{eqnarray*}
\varepsilon^2\int_{\R\times S} dxdy 
\left[\left|\nabla \psi \cdot(1,-Ry\alpha' ) \right|^2 +
\frac{1 }{\varepsilon^2}\left(|\nabla_\perp \psi|^2 - \lambda_0 |\psi|^2 \right) 
-\kappa\frac{|\psi|^2}{\sqrt{x^2+\varepsilon^2 y^2}} \right] ,
\end{eqnarray*} 
for $\psi = \phi \circ f_\alpha^\var \in {\cal H}_0^1(\mathbb R \times S) \subset \LL^2(\mathbb R \times S, \var^2 dxdy)$ and where $R=\left( {\begin{array}{*{20}c}
   0 & -1  \\   1 & 0  \\
\end{array}} \right)$.
Then we perform a division by the global factor~$\varepsilon^2$, so that the full Hilbert space turns to $\LL^2(\R\times S)=\LL^2(\R\times S,dxdy)$.

Hence, we finally obtain the first rescaled family of quad\-rat\-ic forms $b_\kappa^\varepsilon$ we will work with:
\begin{eqnarray}\label{quadForm} b_\kappa^\varepsilon(\psi) &:=&  \int_{\R\times S} dxdy  \\& &\left[\left|\nabla \psi
\cdot(1,-Ry\alpha' ) \right|^2 + \frac{1 }{\varepsilon^2}\left(|\nabla_\perp
\psi|^2-\lambda_0|\psi|^2\right)-\kappa\frac{|\psi|^2}{\sqrt{x^2+\varepsilon^2 y^2}}\right], \nonumber
\end{eqnarray}
and a domain that does not depend on $\kappa,\varepsilon$, that is, $\dom b^\varepsilon_\kappa = \hil^1_0(\R\times S)$ as a subspace of $\LL^2(\mathbb R \times S)$. As usual, write (see Section~9.3 in~\cite{ISTQD}) 
\[
b_\kappa^\varepsilon(\psi)=+\infty\quad\mathrm{ if}\quad \psi\in \LL^2(\R\times S)\setminus \dom b^\varepsilon_\kappa.
\] 
Denote by $H^\varepsilon_\kappa$ the respective operators associated with these forms $b_\kappa^\varepsilon$, and note that all such operators have the same domain $\dom H^\varepsilon_\kappa = \hil^1_0(\R\times S)\cap \hil^2(\R\times S)$, for
which we will actually investigate the limit $\varepsilon\to0$.  The operator $H^\varepsilon_\kappa$ is the relevant rescaled version of $h^\varepsilon_\kappa$ in the repulsive case, and if~$\alpha'(x)=0$ its action is
\[
H^\var_\kappa = -\frac{\partial^2 }{\partial x^2} - \frac{1}{\var^2} \Delta_\perp - \frac{\kappa}{\sqrt{x^2+\var^2 y^2}}-\frac{\lambda_0}{\var^2}.
\]

For the attractive case  we replaced the Coulomb potential by  the regularized version~\eqref{regPot}, with  $0<\delta< 1/2$ (this range of $\delta$ will be necessary for the whole set of results; particularly in the proof of Proposition~\ref{bubu}). The action of the corresponding regularized quad\-rat\-ic form  reads
\begin{eqnarray}\label{quadFormReg}
{a}_\kappa^\var(\psi) &  = &
\int_{\mathbb R \times S} dx dy \\
& & \nonumber
\left[\left|\nabla \psi
\cdot(1,-Ry\alpha' ) \right|^2 + \frac{1}{\var^2} (|\nabla_\perp \psi|^2 - \lambda_0 |\psi|^2)
- \kappa \frac{|\psi|^2}{\sqrt{x^2+ \var^2 y^2}+\var^\delta}  \right],
\end{eqnarray}
$\dom {a}_\kappa^\var = {\cal H}_0^1(\mathbb R \times S)$. This is the second family of rescaled family of quadratic forms we will work with. Note that both forms $b^\var_\kappa$ and ${a}^\var_\kappa$ are positive if~$\kappa<0$, and lower bounded for each~$\var>0$  in case~$\kappa>0$ (by Hardy's Inequality).  Let ${A}_\kappa^\var$ be the self-adjoint operator associated with the quad\-rat\-ic form ${a}_\kappa^\var$; it is our relevant rescaled and regularized version of $h^\varepsilon_\kappa$ in the attractive case, and if~$\alpha'(x)=0$ its action is
\[
{A}^\var_\kappa = -\frac{\partial^2 }{\partial x^2} - \frac{1}{\var^2} \Delta_\perp - \frac{\kappa}{\sqrt{x^2+\var^2 y^2}+\var^\delta}-\frac{\lambda_0}{\var^2}.
\]

\subsection*{Main results}
Now we state the main results of this work, and postpone their proofs to other sections. Let 
\begin{eqnarray*}
  \mathcal D_0 &:=& \left\{w \in \LL^2(\mathbb R): w \in \mathrm{AC}(\mathbb R\setminus\{ 0 \}),
w' \in \LL^2(\mathbb R), w(0^+) = 0 = w(0^-) \right\}\\ &= &\left\{w\in\hil^1(\R): w(0)=0  \right\},
\end{eqnarray*} and denote by $\mathcal D_1$ the collection of elements  $\psi\in\LL^2(\R\times S)$ that can be written in the form $\psi(x,y)=w(x)u_0(y)$ with $w\in \mathcal D_0$. 

The geometric parameter 
\[
C(S) := \int_S | \nabla_\perp u_0 \cdot R y|^2 dy,
\]   was introduced in~\cite{BMT}; it is positive and  depends only on the transverse cross section $S$, and $C(S) > 0$ unless $u_0$ is a radial function. Introduce the {\sc 1D} quad\-rat\-ic form $b^0_\kappa$ by
\begin{equation}\label{quadFormLim}
b^0_\kappa(\psi) :=   \int_{\mathbb R} \left( |w'(x)|^2 + \alpha'(x)^2 C(S)\; |w(x)|^2 - \kappa\,
\frac{|w(x)|^2}{|x|} \right) \, dx
\end{equation} if $ \psi = w u_0\in \mathcal D_1$, and we will alternatively write simply $b^0_\kappa(w)$. Note that,  by means of projections, the form $b^0_\kappa$ may be seen as acting either  with domain the subspaces $\mathcal D_0$ (in $\R$) or with domain $\mathcal D_1$ (in $\R\times S$), and this is a key step for the {\em dimensional reduction} we study here.

 The self-adjoint operator in $\R$ associated with $b^0_\kappa$ is (recall that $H_D$ is defined in Appendix~\ref{1dHatom})
\[
(H^0_\kappa w)(x) = (H_Dw)(x)+ \alpha'(x)^2 C(S)w(x), 
\]acting on a dense subset of $\mathcal D_0$ in $\LL^2(\R)$, and with Dirichlet boundary condition; a proof of this fact appears in Appendix~\ref{appHkappa0}. Recall that $\alpha'$ is a bounded function, so that the respective term brings no issue to the domain of the operator $H_\kappa^0$, that is, $\dom H_\kappa^0=\dom H_D$ (see equation~\eqref{DirichletHamiltonian}). Note that the operator $H^0_\kappa$ may, correspondingly to the forms, be seen as acting in subspaces of either $\mathcal D_0$ (in $\R$) or $\mathcal D_1$ (in $\R\times S$).

\begin{teor2}\label{teorAttractive}[Attractive Case]
Assume that $\kappa>0$ and $0< \delta < 1/2$.
Then, ${A}^\varepsilon_\kappa$ converges to $H^0_\kappa$ in the norm resolvent sense as $\varepsilon\to0$. More precisely,
\[
\left\|    \left({A}_\kappa^\var -  i {\bf 1} \right)^{-1}    
- \left[\left(H^0_\kappa -  i  {\bf 1} \right)^{-1} \oplus 0 \right]\right\| \to 0
\]
as $\var \to 0$, where $0$ is the null operator on the subspace $\{ w u_0: w \in \LL^2(\mathbb R)\}^\perp$.
\end{teor2}

\begin{teor2}\label{teorRepulsive}[Repulsive Case]
If $\kappa<0$, then the family of operators $H^\varepsilon_\kappa$ converges to $H^0_\kappa$ in the strong resolvent sense as $\varepsilon\to0$. More precisely, for all $\vartheta\in\LL^2(\R\times S)$ one has
\[
\left\| (H^\var_\kappa - i \Id)^{-1} \vartheta - \left[(H^0_\kappa - i \Id)^{-1} \oplus 0\right] \vartheta\right\| \to 0
\]
as $\var \to 0$, where $0$ is the null operator on the subspace 
$\{w u_0: w \in \LL^2(\mathbb R)\}^\perp$. 
\end{teor2}

\begin{obs2}
Note that if there is no rotation of the cross section $S$, i.e., if $\alpha$ is a constant function, then the limit operator $H^0_\kappa$ after confinement is always~$H_D$ and it does not depend on the shape of~$S$.
\end{obs2}

\section{Attractive Coulomb potential}\label{sectAttractive}

 Recall that it is enough to get a convergence for just one point in the common resolvent set of the involved operators; here, sometimes we write them in terms of convergence of quad\-rat\-ic forms.

An important part of our strategy is the introduction of intermediary operators through suitable   modifications of the  quad\-rat\-ic forms. First, just to simplify expressions, we assume that $\alpha'(x)=0$ for all $x\in\R$. Second, we add a suitable constant potential term $c / \var^\delta \|\psi\|^2$, for some $c>0$, and the restriction $0<\delta< 1/2$ turns the involved quad\-rat\-ic forms into 
positive ones, so that  the modification $a^\var_\kappa$ of quad\-rat\-ic form~\eqref{quadFormReg}   reads
\begin{equation}\label{quadFormLimMod}
\dot a_\kappa^\var(\psi) = {a}_\kappa^\var (\psi) + \frac{c}{\var^\delta} \int_{\R\times S}  |\psi|^2\,dx dy  ,
\end{equation}
$\dom{\dot a}_\kappa^\var = {\cal H}_0^1(\mathbb R \times S)$.

Our first result stated in this section says that the convergence in the norm resolvent sense of the sequence of operators associated with these modified forms
can only occur on a specific subspace.
More precisely, our analysis is restricted to the sequence of one dimensional quad\-rat\-ic forms
\begin{equation}\label{ppp3}
t_\kappa^\var(w) ={\dot a}_\kappa^\var(w u_0) =
\int_\mathbb R 
\left[|w'|^2 + V_\kappa^\var(x) |w|^2 + \frac{c}{\var^\delta} |w|^2  \right] dx,
\end{equation}
with
$$V_\kappa^\var(x) := - \kappa \int_S \frac{|u_0|^2}{\sqrt{x^2+\var^2 y^2}+\var^\delta}dy,$$
and $\dom t_\kappa^\var = {\cal H}^1(\mathbb R)$.
Observe that, among other properties that we will use ahead,
$V_\kappa^\var(x) \rightarrow - \kappa/|x|$ a.e.[$x$] as $\varepsilon\to0$.

Let ${\dot A}_\kappa^\var$ be the self-adjoint operator associated with the quad\-rat\-ic form $\dot a_\kappa^\var$
and  $T_\kappa^\var$ the self-adjoint operator associated with  $t_\kappa^\var$. We will prove the following results:

\begin{prop2}\label{bubu}
Assume that $\kappa\ne 0$ and $0< \delta < 1/2$.
Then, there exists a number $C> 0$ so that, for $\var>0$ small enough,
\[
\big\|    \left({\dot A}_\kappa^\var\right)^{-1}    
- \left[\left(T_\kappa^\var\right)^{-1} \oplus 0\right] \big\| \leq   C\, \var^{1+\delta/2},
\]
where $0$ is the null operator on the subspace $\{ w u_0: w \in \LL^2(\mathbb R)\}^\perp$.
\end{prop2}

\begin{prop2}\label{thidrresolvent}
Assume that $\kappa\ne 0$ and $0< \delta < 1/2$.
Then, there exists a number $D> 0$ so that, for $\var>0$ small enough,
\[
\left\|    \left({\dot A}_\kappa^\var - \left(\frac{c}{\var^\delta} + i\right) {\bf 1} \right)^{-1}    
- \left[\left(T_\kappa^\var  - \left( \frac{c}{\var^\delta} + i \right) {\bf 1} \right)^{-1} \oplus 0 \right]\right\| \leq   D \, \var^{1-3\delta/2},
\]
where $0$ is the null operator on the subspace $\{ w u_0: w \in \LL^2(\mathbb R)\}^\perp$.
\end{prop2}

The next result relates the convergence of~$T_\kappa^\var$, in the norm resolvent sense, to the self-adjoint extension  of the 1D hydrogen atom with
Dirichlet condition at the origin.

\begin{prop2}\label{hihi} Assume that $\kappa> 0$. In the space $\LL^2(\R)$, the family of operators
$(T_\kappa^\var -  c / \var^\delta {\bf 1})$ converges in the norm resolvent sense to~$H_D$ as $\var \rightarrow 0$.
\end{prop2}

Since ${A}^\var_\kappa = {\dot A}^\var_\kappa -c/\var^\delta$, by combining Propositions~\ref{thidrresolvent} and~\ref{hihi}  we obtain  Theorem~\ref{teorAttractive}. 
In what follows we present the proofs of the above propositions.

\subsection{Reduction of dimension}\label{rdd}

First, we are going to make explicit the constant $c >0$ that was introduced
in the quad\-rat\-ic forms ${\dot a}_\kappa^\var$; note that even in the repulsive case it will be convenient to  keep the term $c / \var^\delta |\psi|^2$ in the quadratic forms. Observe that
\[
-\frac{\kappa}{|(x, \var y)|+\var^{\delta}} + \frac{\kappa}{\var^\delta} \geq 0.
\]
Thus, we choose $c> \kappa > 0$ so that
\[
-\frac{\kappa}{|(x, \var y)|+\var^{\delta}} + \frac{c}{\var^\delta} \geq \frac{d}{\var^{\delta}},
\]
for some $d> 0$.
This guarantees that
\[
\dot a_\kappa^\var(\psi) \geq \frac{d}{\var^{\delta}} \int_{\mathbb R \times S} |\psi|^2 dxdy,
\qquad \forall \psi \in{\cal H}_0^1(\mathbb R \times S).
\]
This inequality will be important ahead.

Let ${\cal L}$ be a subspace of  $\LL^2(\mathbb R \times S)$
generated by all functions of the form 
$$\psi_w (x,y)=  w(x) u_0(y), 
\quad  w \in \LL^2(\mathbb R).$$
Recall that $u_0$ is the normalized eigenfunction associated with the eigenvalue  $\lambda_0$ of the
Laplacian in 
${\cal H}_0^1(S)$.
We consider the orthogonal decomposition $\LL^2(\mathbb R \times S) = {\cal L} \oplus {\cal L}^\perp.$
Thus, if $\psi \in  \LL^2(\mathbb R \times S)$, we can write
$$\psi = \psi_w + \eta, \quad w \in \LL^2(\mathbb R), \quad  \eta \in {\cal L}^\perp.$$

We take now
$\eta \in {\cal L}^\perp \cap {\cal H}_0^1(\mathbb R \times S)$. Then,
$$\int_S  \eta(x, y) u_0(y) dy = 0 \quad  \hbox{and}
\quad 
\int_S  \eta'(x, y) u_0(y) dy = 0 \quad  \hbox{a.e.[x]},$$
where $\eta'(x,y)$ denotes the derivative of $\eta$ with respect to $x$.
Using the Green Identities, we can also see that 
$$\int_S  \langle \nabla_\perp \eta, \nabla_\perp u_0 \rangle dy = 0,
\quad  \hbox{a.e.[x]}.$$

Let $\lambda_1$ denote the second eigenvalue of the Laplacian on
${\mathcal H}_0^1(S)$ and recall that $\lambda_1>\lambda_0$. Since $\eta \bot u_0$ in $\LL^2(S)$, we have
$$\int_S  |\nabla_\perp \eta|^2  dy \geq \lambda_1 \int_S |\eta|^2 dy , \quad  \hbox{a.e.[x]}.$$
Consequently
$$\int_S \left( |\nabla_\perp \eta|^2 - \lambda_0 |\eta|^2 \right) dy 
\geq (\lambda_1 - \lambda_0) \int_S  |\eta|^2 dy, \quad  \hbox{a.e.[x]}.$$

In the particular case of a function 
$w \in {\cal H}^1(\mathbb R)$ one has $\psi_w \in {\cal H}_0^1(\mathbb R \times S)$. Therefore,
for $\psi \in {\cal H}_0^1(\mathbb R \times S)$, we write
$\psi = \psi_w + \eta$ with $w \in {\cal H}^1(\mathbb R)$ and $\eta \in {\cal L}^\perp \cap {\cal H}_0^1(\mathbb R \times S)$.

As already previously mentioned, we are going  to analyze the sequence
${\dot a}_\kappa^\var(\psi)$ with
$\dom{\dot a}_\kappa^\var = {\cal H}_0^1(\mathbb R \times S)$ (see just before Eq.~\eqref{ppp3}), and recall that 
${\dot A}_\kappa^\var$ denotes the self-adjoint operator associated with~${\dot a}_\kappa^\var$.
For $\psi = \psi_w$, $w \in {\cal H}^1(\mathbb R)$, we have
\[
{\dot a}_\kappa^\var(\psi_w) =
\int_{\mathbb R \times S} dx dy \left[|w'|^2 |u_0|^2 - \kappa \frac{|w|^2|u_0|^2}{\sqrt{x^2+\var^2 y^2}+\var^\delta} 
+ \frac{c}{\var^\delta} |w|^2 |u_0|^2 \right]. 
\]
Recall that  
$T_\kappa^\var$ denotes the self-adjoint operator associated with~$t_\kappa^\var$ (see Eq.~\eqref{ppp3}) and $\dom T_\kappa^\var = {\cal H}^2(\mathbb R)$.

\subsubsection{Proof of  Proposition~\ref{bubu}}
\noindent
For the sake of simplicity, in this proof we  suppose the involved vectors are real; all calculations can be easily adapted for complex vectors. As already mentioned, for $\psi \in {\cal H}_0^1(\mathbb R \times S)$,  we write
$\psi(x, y) = \psi_w(x, y) + \eta(x, y)$ with $w \in {\cal H}^1(\mathbb R)$ and $\eta \in {\cal H}_0^1(\mathbb R \times S) \cap
{\cal L}^\perp$.
Thus,
$$\dot a_\kappa^\var(\psi) = t_\kappa^\var(w) + \dot a_\kappa^\var(\eta) + 2 m_\kappa^\var(\psi_w, \eta),$$
with $t_\kappa^\var(w)$  given by \eqref{ppp3},
$$\dot a_\kappa^\var(\eta) = \int_{\mathbb R \times S}
\left[|\eta'|^2 + \frac{1}{\var^2}\left(|\nabla_\perp \eta|^2 - \lambda_0|\eta|^2\right)
- \kappa \frac{|\eta|^2}{\sqrt{x^2+\var^2 y^2}+\var^\delta} + \frac{c}{\var^\delta} |\eta|^2 \right] dx dy$$
and
$$m_\kappa^\var(\psi_w, \eta) : = - \int_{\mathbb R \times S}
\kappa \left( \frac{w u_0 \eta}{\sqrt{x^2+ \var^2 y^2}+\var^\delta} \right) dx dy.$$

We are going to check that there are $c_0>0$ and functions $0\le q(\var),$ $0\le p(\var)$ and $c(\var)$ so that $t_\var(w)$, ${\dot a}_\kappa^\var(\eta)$ and $m_\var(\psi_w, \eta)$ satisfy
the following conditions:
\begin{equation}\label{pipoca001}
t_\var(w) \geq c(\var) \|w \|_{\LL^2(\mathbb R)}^2, \quad \forall  w \in {\cal H}^1(\mathbb R), \quad c(\var) \geq c_0 > 0;
\end{equation}
\begin{equation}\label{pipoca002}
{\dot a}^\var_\kappa(\eta) \geq p(\var) \|\eta \|_{\LL^2(\mathbb R \times S)}^2, \quad
\forall \eta \in {\cal H}_0^1(\mathbb R \times S) \cap
{\cal L}^\perp;
\end{equation}
\begin{equation}\label{pipoca003}
|m_\var(\psi_w, \eta)|^2 \leq q(\var)^2\; t_\var(w)\;{\dot a}^\var_\kappa(\eta), \quad \forall \psi \in {\cal H}_0^1(\mathbb R \times S);
\end{equation}
and with
\begin{equation}
p(\var) \to \infty,\; c(\var) = O (p(\var)),\; q(\var) \to 0 \quad \hbox{as} \quad \var \to 0.
\end{equation}
Thus,  Theorem~2 in~\cite{dOVSantiago} (which is in fact just an abstract reformulation of Proposition~3.1 in~\cite{Sol}) guarantees
that, for $\var > 0$ small enough, 
\begin{equation}\label{pipoca004}
\left\| ({\dot A}_\kappa^\var)^{-1} - (T_\kappa^\var)^{-1} \oplus 0 \right\| \leq  p(\var)^{-1} + C'  q(\var)\, c(\var)^{-1},
\end{equation}
for some $C' > 0$.

By construction,
$$t_\kappa^\var(w) \geq \frac{d}{\var^\delta} \int_\mathbb R |w|^2dx.$$
We take $c(\var): = d /\var^\delta$. Since $d > 0$, it follows that $c(\var) \to \infty$ as $\var \to 0$.
Now, since
$\eta \in {\cal H}_0^1(\mathbb R \times S) \cap {\cal L}^\perp$,
\begin{eqnarray*}
{\dot a}_\kappa^\var(\eta) & \geq &
\int_{\mathbb R \times S} \frac{1}{\var^2} \left(|\nabla_\perp \eta|^2 - \lambda_0 |\eta|^2\right) dx dy\\
& \geq  & \frac{\lambda_1 - \lambda_0}{\var^2} \int_{\mathbb R \times S}  |\eta|^2 dx dy.
\end{eqnarray*}
Since $\lambda_1 - \lambda_0 >0$, we take $p\displaystyle(\var) = \frac{\lambda_1 - \lambda_0}{\var^2} \to \infty$, as $\var \to 0$ and note that $c(\var) = {\rm O}(p(\var))$.

Finally,   there exists a number $C''> 0$ so that, 
\begin{eqnarray*}
|m_\kappa^\var(\psi_w, \eta)| & \leq &
|\kappa| \left| \int_{\mathbb R \times S} \frac{\psi_w \eta}{\sqrt{x^2+\var^2 y^2}+\var^\delta} dx dy \right| \\
& \leq & 
\frac{|\kappa|}{\var^\delta} \int_{\mathbb R \times S} |\psi_w| |\eta| dx dy \\
& \leq &
\frac{|\kappa|}{\var^\delta} \left( \int_{\mathbb R \times S} |\psi_w|^2 dx dy  \right)^{1/2}
\left( \int_{\mathbb R \times S} |\eta|^2 dx dy  \right)^{1/2}\\
& = &
\frac{|\kappa|}{\var^\delta} \left( \int_\mathbb R |w|^2 dx    \right)^{1/2}
\left( \int_{\mathbb R \times S} |\eta|^2 dx dy  \right)^{1/2} \\
& \leq &
\frac{C''}{\var^\delta} \, \var^{\delta/2} \, \var \, (t_\kappa^\var(w))^{1/2} ({\dot a}_\kappa^\var(\eta))^{1/2} \\
& \leq &
q(\var)\, (t_\kappa^\var(w))^{1/2} \,({\dot a}_\kappa^\var(\eta))^{1/2},
\end{eqnarray*}
where
$q(\var) :=   C'' \, \var^{1-\delta/2} \to 0$ as $\var \to 0$
(recall that, by hypothesis, $0 < \delta < 1/2$). 

The conditions (\ref{pipoca001}), (\ref{pipoca002}) and (\ref{pipoca003}) are then satisfied and  we  conclude that inequality~\eqref{pipoca004} holds true, from which one finds, for $\var > 0$ small enough, 
$$\left\|({\dot A}_\kappa^\var)^{-1} - \left[(T_\kappa^\var)^{-1} \oplus 0\right]\right\| \leq C \var^{1+\delta/2},$$
for some $C> 0$. The proof of the proposition is complete.

\subsubsection{Proof of  Proposition~\ref{thidrresolvent}}
We shall make use of  the third resolvent identity~\cite{ISTQD}, that is, if~$S$ and~$T$ are linear operators and $z,z_0$ are common elements of the resolvent sets of both~$S$ and~$T$, then \[
R_z(T)-R_{z}(S) = \left(\Id+(z-z_0)R_z(T)  \right)\left[R_{z_0}(T)-R_{z_0}(S)  \right]  \left(\Id+(z-z_0)R_z(S) 
\right).
\]

If for a moment we denote $\xi=\left({c}{\var^{-\delta}} + i\right)$, by such identity one has
\begin{eqnarray*}\Big({\dot A}_\kappa^\var - \xi {\bf 1} \Big)^{-1}    
&-& \Big[\Big(T_\kappa^\var  - \xi {\bf 1} \Big)^{-1} \oplus 0   \Big] \\ &=& 
\left[ {\bf 1} + \xi \left( {\dot A}_\kappa^\var - \xi {\bf 1} \right)^{-1}\right]
\\ & &
\left[ ({\dot A}_\kappa^\var)^{-1}   - (T_\kappa^\var)^{-1} \oplus 0\right]
\left[   
{\bf 1}  + \xi \left[\left( T_\kappa^\var - \xi {\bf 1} \right)^{-1}
\oplus 0\right]
\right].
\end{eqnarray*}
Thus, since for all~$\var>0$
$$\left\| \left({\dot A}_\kappa^\var - \left({c}{\var^{-\delta}} + i\right) {\bf 1} \right)^{-1} \right\| \leq 1
\quad \hbox{and} \quad
\left\| \left(T_\kappa^\var - \left( {c}{\var^{-\delta}} + i\right) {\bf 1} \right)^{-1}  \oplus 0 \right\| \leq 1,$$
 by Proposition~\ref{bubu}, it is found that
\begin{eqnarray*}
\big\| 
\big({\dot A}_\kappa^\var - \big( {c}{\var^{-\delta}} + i  \big)  {\bf 1} \big)^{-1}    
&-& \big(T_\kappa^\var  - \big( {c}{\var^{-\delta}} + i \big) {\bf 1}  \big)^{-1} \oplus 0  \big\| \\
&\leq& 
\left( 1 + \sqrt{c^2/\var^{2\delta}+ 1} \right) \var^{1+\delta/2}
\left( 1 + \sqrt{c^2/\var^{2\delta} + 1} \right)\\
&\leq& D \, \var^{1-3/2\delta},
\end{eqnarray*}
for some~$D > 0$.  This completes the proof of the proposition.

\subsubsection{Proof of  Proposition~\ref{hihi}}

For each $\var > 0$, write
\begin{equation}\label{ppp1}
(H_\var w)(x) := - w''(x) + W_\var(x) w(x), \quad
\dom H_\var = {\cal H}^2(\mathbb R),
\end{equation}
where $W_\var$ is a potential so that the above operator  is well defined and
 self-adjoint.
We assume that the limit $W_0(x):= \lim_{\var \to 0} W_\var(x)$  exists a.e.,
 and also that 
$W_0(x)$ is a potential with a singularity at the origin so that the operator
$$(H_0 w)(x) := - w''(x) + W_0(x) w(x),
$$
with $\dom H_0 = \{{\cal H}^2(\mathbb R \backslash \{0\}): w(0^\pm)=0\}$
is self-adjoint.

Ahead we will make use of the following result proven in~\cite{Klaus}:

\begin{teor2}\label{ppp2}
{\rm
Let $H_\var$ be as in \eqref{ppp1} and  $W_\var(x)$ satisfying:
\begin{itemize}
\item[(i)] there exists a number $ g > 0$ so that
$$|W_\var(x)| \leq g \left(\frac{1}{|x|^\gamma} + \frac{1}{|x|^\beta}\right), \quad  
0 < \var \leq \var_0,$$
where $\beta > 0$ and $0 < \gamma < 2$;
\item[(ii)] $W_\var(x) \in \LL_{{\rm loc}}^1(\mathbb R)$, $0 < \var \leq \var_0$;
\item[(iii)] $W_\var(x) \to W_0(x)$ a.e. as  $\var \to 0$;
\item[(iv)] $\dis \int_b^{-b} W_\var(x) dx \to - \infty$ as $\var \to 0$ 
for some $b> 0$;
\item[(v)]
$\dis \sup_{\var \in (0, \var_0)} \frac{\int_b^{-b} |W_\var(x)| dx }{\left|\int_b^{-b} W_\var(x) dx \right|} < \infty$.
\end{itemize}
Then, $H_\var$ converges to $H_0$ in the norm resolvent sense as $\var \to 0$.
}
\end{teor2}

To prove Proposition~\ref{hihi}, we are going to show that 
\[
\dis V_\kappa^\var(x)  = - \kappa \int_S \frac{|u_0|^2}{\sqrt{x^2+\var^2 y^2}+\var^\delta} dy
\] 
satisfies the conditions (i)-(v) in Theorem~\ref{ppp2}. 
To show~(i) it is enough to observe that
\[|V_\kappa^\var(x)|  \leq  \kappa \int_S \frac{|u_0|^2}{|x|+\var^\delta} dy 
= \frac{\kappa}{|x|+\var^\delta}  \leq  \frac{\kappa}{|x|},
\]
and then consider $g = \kappa / 2$ and $\gamma = \beta = 1.$
It is easy to see that $V_\kappa^\var \in \LL_{{\rm loc}}^1(\mathbb R)$ and so (ii) is satisfied.
To prove~(iii), we apply the Dominated Convergence Theorem and obtain 
\begin{eqnarray*}
\lim_{\var \to 0} \int_S \frac{|u_0|^2}{\sqrt{x^2 + \var^2 y^2}+\var^\delta} dy
& = &
\int_S \lim_{\var \to 0}  \frac{|u_0|^2}{\sqrt{x^2 + \var^2 y^2}+\var^\delta} dy\\
& = &
\int_S \frac{|u_0|^2}{|x|} dy = \frac{1}{|x|},
\end{eqnarray*}
i.e., $\dis \lim_{\var \to 0} V_\kappa^\var(x) \to - \frac{\kappa}{|x|} = V_\kappa(x).$

Now we are going to check~(iv). Since $S$ is a bounded region, there exists a 
number $K> 0$ so that 
$y^2 \leq K$. Thus,
\begin{eqnarray*}
\int_{-1}^{1} V_\kappa^\var(x) dx & \leq & - \kappa \int_{-1}^1 \int_S \frac{|u_0|^2}{\sqrt{x^2+\var^2 K}+\var^\delta} dx dy \\
& = & - \kappa
\int_{-1}^1 \frac{1}{\sqrt{x^2+\var^2 K }+\var^\delta}  dx.
\end{eqnarray*}
Consequently,
$$\lim_{\var \to 0} \int_{-1}^{1} V_\kappa^\var(x) dx = - \infty.$$
Condition (v) is clearly satisfied.
Hence, the proof of Proposition~\ref{hihi} follows as an application of Theorem~\ref{ppp2}.

\begin{obs2}
Note that item~(iv) of Theorem~\ref{ppp2} does not apply to $T_\kappa^\var$ if $\kappa < 0$. 
\end{obs2}

\subsection{A condition without regularization}
Based on Theorem~\ref{teorAttractive}, a natural question in the attractive case is about the approximation of $R_i({A}^\var_\kappa)\vartheta$ by $R_i(H^\var_\kappa)\vartheta$, so that, after combining  with other propositions, one could conclude a convergence of~$H^\var_\kappa$ to~$H_D$, that is, a convergence without regularization. We discuss a possible suitable approximation and it makes clear the connection with bounded energy values (which vaguely reminds us of arguments in~\cite{deO_PLA}).

Given $\vartheta\in \LL^2(\R\times S)$, for each $\var>0$ there is a unique $\psi^\var\in \dom H^\var_\kappa$ with $(H^\var_\kappa -i\Id)\psi^\var = \vartheta$; note that $\|\psi^\var\|=\|R_i(H^\var_\kappa)\vartheta\|\le \|\vartheta\|$, that is, there is a uniform upper bound for the norm of~$\psi^\var$. 

By the second resolvent identity,
\begin{eqnarray}
\left\| R_i({A}^\var_\kappa)\vartheta - R_i(H^\var_\kappa)\vartheta  \right\|^2  \nonumber &=& \left\| R_i({A}^\var_\kappa)\left( H^\var_\kappa-{A}^\var_\kappa \right) R_i(H^\var_\kappa)\vartheta  \right\|^2 \\ \nonumber &\le & \left\|  \left(\frac{\kappa}{|(x,\var y)|+\var^\delta} - \frac{\kappa}{|(x,\var y)|}  \right) R_i(H^\var_\kappa)\vartheta \right\|^2 \\ &=& \label{eqQeps}
\kappa^2 \var^{2\delta} \left \|  \frac{\psi^\var(x,\var y)}{|(x,\var y)|\left(|(x,\var y)|+\var^\delta\right)}    \right\|^2 := Q^\var,
\end{eqnarray}
and after the successive change of variables $r^2=y_1^2+y_2^2$, $z=\var r$ and $u=x^2+z^2$, the right hand side $Q^\var= Q^\var(\psi^\var)$ of~\eqref{eqQeps} reads
\[
Q^\var = \kappa^2 \var^{2(\delta-1)} \int_\R dx \int_{x^2}^{x^2+\var^2} du\, \frac{|\psi^\var(x,u)|^2}{\left(u+\var^\delta \sqrt{u} \right)^2}\,,
\]
and the required convergence is directly related to the vanishing of $Q^\var$ as~$\var\to0$. This quantity represents quantitatively the proclaimed energy control as~$\var\to0$. We mention  two different possibilities one is able to estimate $Q^\var$:
\begin{enumerate}
\item Assume that for all $\var>0$ small enough one has $|\psi^\var(x,u)|\le |w(x)|\,u^p$, for some $w\in\LL^2(\R)$ and $3/4<p<1$. Note that this is a special case of Dirichlet condition at the origin for the 3D wavefunctions. Thus,
\begin{eqnarray*}
Q^\var &=& \kappa^2 \var^{2(\delta-1)} \int_\R dx\, |w(x)|^2 \int_{x^2}^{x^2+\var^2}du \left( \frac{u^p}{u+\var^\delta\,\sqrt{u}}  \right)^2
 \\ &\le&  \kappa^2 \var^{2(\delta-1)} \int_\R dx\, |w(x)|^2 \int_{x^2}^{x^2+\var^2}du\; u^{2(p-1)}
  \\ &\le&  \frac{\kappa^2}{2p-1} \var^{2(\delta-1)} \int_\R dx\, |w(x)|^2\left( (x^2+\var^2)^{2p-1}-(x^2)^{2p-1}\right) 
  \\ &\le&  \frac{\kappa^2}{2p-1} \var^{2(\delta-1)} \int_\R dx\, |w(x)|^2  \var^{4p-2}
  \\  &=&  \frac{\kappa^2}{2p-1} \var^{2\delta+4p-4} \|w\|^2.
\end{eqnarray*}
Hence, given such~$p$, by choosing $\delta$ so that  $2-2p <\delta <1/2$ one has 
\[
\left\| R_i({A}^\var_\kappa)\vartheta - R_i(H^\var_\kappa)\vartheta  \right\| \to 0,
\] since $Q^\var\to 0$, as $\var\to0$.

\item Suppose now that $|\psi^\var(x,u)|\ge M>0$ in a neighbourhood of the origin (so, far from the Dirichlet condition at the origin), which for simplicity we assume this inequality holds true in the set $(x,u)\in  [0,1/2]\times [0,1/4+\var^2]$. Since in this set $u\le\sqrt{u}$, one has, for some $D>0$ that does not depend on~$\var$,
\begin{eqnarray*}
Q^\var &\ge& \kappa^2 M^2 \var^{2(\delta-1)}  \int_0^{1/2} dx \int_{x^2}^{x^2+\var^2} 
\frac{du}{\left((1+\var^\delta)\sqrt{u}\right)^2}
\\ &\ge& D \var^{2(\delta-1)} \int_0^{1/2} \ln \left( 1+\frac{\var^2}{x^2} \right)
\\ &=& D \left( 2\var^{2\delta-1} \arctan\left(\frac{1}{2\var}\right) + \frac12 \var^{2(\delta-1)} \ln(1+4\var^2) \right)
\\ &\sim& D \var^{2(\delta-1)}\to\infty
\end{eqnarray*}as $\var\to0$. Hence, the convergence is not expected to hold in this case. Note that the same argument shows that $\|{A}^\var_\kappa \psi^\var - H^\var_\kappa\psi^\var\|\to\infty$ as~$\var\to0$.
\end{enumerate}

\

\section{Repulsive Coulomb potential}\label{sectRepulsive}
In this section we consider the repulsive Coulomb potential $V_C$ introduced in equation \eqref{potVC}, i.e., $\kappa<0$; we aim at proving Theorem~\ref{teorRepulsive}, and this  will be attained by showing that statement (j) in Appendix~\ref{appendixGammaSR} holds true. Recall that $u_0$ is the positive normalized eigenfunction corresponding to the lowest (simple) eigenvalue of $-\Delta_\perp$ in $S$. First we show that  $b^\varepsilon_\kappa$ strongly $\Gamma$-converges to $b^0_\kappa$. We begin with some basic properties and remarks:

\begin{enumerate} \item \label{proper1} Let $\psi \in{\mathcal H}_0^1(\mathbb R \times S)$; then, for $x$-a.s.\   one has
\[
\frac{1}{\varepsilon^2} \int_S \left( | \nabla_\perp \psi|^2 - \lambda_0 |\psi|^2\right)\, dy \geq 0.
\]
\item \label{proper2} The Dirichlet boundary condition at the border of~$S$ implies that 
\[
\displaystyle\int_S \left( u_0 \nabla_\perp u_0 \cdot R y \right) dy = 0.
\]
\item \label{proper3} Let $\psi_\varepsilon \to  \psi$ in $\LL^2(\mathbb R \times S)$, as $\varepsilon\to0$, and so that $(b^\varepsilon_\kappa(\psi_\varepsilon))_\varepsilon$ is a bounded sequence in $\R$. Then $(\psi_\varepsilon')_\varepsilon$ and $(\nabla_\perp \psi_\varepsilon)_\varepsilon$ are bounded sequences in $\LL^2(\mathbb R \times S)$.
Therefore, by taking subsequences if necessary, it follows that $ \psi_\varepsilon' \rightharpoonup \psi'$ and $\nabla_\perp \psi_\varepsilon \rightharpoonup \nabla_\perp \psi$ weakly in $\LL^2(\mathbb R \times S)$ and $\psi \in{\mathcal H}_0^1(\mathbb R \times S)$.

Indeed, first note that there exists $C>0$ so that
\[
\limsup_{\varepsilon \to  0} \int_{\mathbb R} \int_S 
|\psi_\varepsilon' + \nabla_\perp \psi_\varepsilon \cdot R y \alpha'|^2 \, dydx
\leq \limsup_{\varepsilon \to  0}  b^\varepsilon_\kappa (\psi_\varepsilon) \leq C
\]
and
\begin{eqnarray*}
& &\limsup_{\varepsilon \to  0}  \int_{\mathbb R} \int_S |\nabla_\perp \psi_\varepsilon|^2 \, dydx \\
& = & 
\limsup_{\varepsilon \to  0} \left( 
\int_{\mathbb R} \int_S \left(|\nabla_\perp \psi_\varepsilon|^2 - \lambda_0 |\psi_\varepsilon|^2 \right) dydx
+ \lambda_0 \int_{\mathbb R} \int_S |\psi_\varepsilon|^2 \, dydx
\right) \\
& \leq &
\limsup_{\varepsilon \to  0} C \varepsilon^2
+ \lambda_0 \limsup_{\varepsilon \to  0}   \int_{\mathbb R} \int_S |\psi_\varepsilon|^2 \, dydx
< \infty.
\end{eqnarray*}
Since $\alpha' \in \LL^\infty(\mathbb R)$, it follows that
$(\psi'_\varepsilon, \nabla_\perp \psi_\varepsilon )$ is a bounded sequence in $\LL^2(\mathbb R \times S)$.
Hence, $(\psi_\varepsilon)_\varepsilon$ is a bounded sequence in ${\mathcal H}_0^1(\mathbb R \times S)$.

\item\label{prop4} Let $\psi_\varepsilon \to  \psi$ in $\LL^2(\mathbb R \times S)$ so that 
$\lim_{\varepsilon \to  0} b^\varepsilon_\kappa(\psi_\varepsilon) < \infty$.
Then, we can write $\psi(x,y)=w(x)u_0(y)$ with $w \in{\mathcal H}^1(\mathbb R)$.

Indeed, by the previous item, $\nabla_\perp \psi_\varepsilon \rightharpoonup \nabla_\perp \psi$  in
$\LL^2(\mathbb R \times S)$. Thus,
\begin{eqnarray*}
\int_{\mathbb R} \int_S |\nabla_\perp \psi|^2 \, dydx & \leq& 
\liminf_{\varepsilon \to  0} \int_{\mathbb R} \int_S |\nabla_\perp \psi_\varepsilon|^2 \, dydx\\
&\leq&  \lambda_0 \limsup_{\varepsilon \to  0}  \int_{\mathbb R} \int_S |\psi_\varepsilon|^2 \, dydx
\\ &=& \lambda_0 \int_{\mathbb R} \int_S |\psi|^2 \, dydx,
\end{eqnarray*}
and so 
\begin{equation}\label{intNula}
\int_{\mathbb R} \int_S \left(|\nabla_\perp \psi|^2 - \lambda_0 |\psi|^2\right)\, dydx = 0.
\end{equation}

Since by item \ref{proper1}.\ above one has 
\[
0\le f(x) := \int_S \left(|\nabla_\perp \psi(x,y)|^2 - \lambda_0 |\psi(x,y)|^2\right) dy,
\]
the  equality \eqref{intNula} implies that    $f=0$ $x$-a.s.,  and so $\psi(x, \cdot)$ is an eigenvector associated with the (simple) eigenvalue $\lambda_0$, that is, $\psi(x, \cdot)$ is proportional to $u_0(\cdot)$. Hence, we can write, $\psi(x,y) = w(x) u_0(y)$ on putting $w \in{\mathcal H}^1(\mathbb R)$  (since $\psi \in{\mathcal H}_0^1(\mathbb R \times S)$).

\end{enumerate}

By keeping in mind the above remarks, we divide the proof of Theorem~\ref{teorRepulsive} in three steps. Note that, without loss of generality, we may assume that $\kappa=-1$.

\

\noindent {\bf Step 1.} Let $\psi_\varepsilon \to  \psi$ in $\LL^2(\mathbb R \times S)$. 

If $\liminf_{\varepsilon \to  0} b^\varepsilon_\kappa(\psi_\varepsilon) = \infty$, then automatically 
$\liminf_{\varepsilon \to  0} b^\varepsilon_\kappa(\psi_\varepsilon) \geq b^0_\kappa(\psi)$.
Suppose then that  $\liminf_{\varepsilon \to  0} b^\varepsilon_\kappa(\psi_\varepsilon) < \infty$.
By taking a subsequence (if necessary), one may suppose that 
\[
\liminf_{\varepsilon \to  0} b^\varepsilon_\kappa(\psi_\varepsilon) = \lim_{\varepsilon \to  0} b^\varepsilon_\kappa(\psi_\varepsilon) < \infty.
\]
By properties \ref{proper3}.\ and~\ref{prop4}.\ above, it follows that
$\psi \in{\mathcal H}_0^1(\mathbb R \times S)$, 
$ \psi_\varepsilon' \rightharpoonup \psi'$ and $\nabla_\perp \psi_\varepsilon \rightharpoonup \nabla_\perp \psi$ weakly in
$\LL^2(\mathbb R \times S)$, and so one may write 
$\psi(x, y)=w(x)u_0(y)$ for some $w \in{\mathcal H}^1(\mathbb R)$.
Particularly, since $\alpha \in \LL^\infty(\mathbb R)$ one has
$$ \psi'_\varepsilon + \nabla_\perp \psi_\varepsilon \cdot R y \alpha' \rightharpoonup 
\psi' + \nabla_\perp \psi \cdot R y \alpha'.$$

Again, by passing to a subsequence if necessary, since
$\psi_\varepsilon \to  \psi$ in $\LL^2(\mathbb R \times S)$, one has,  $(x,y)$-{a.e.}
\[
\psi_\varepsilon(x,y) \to  w(x)u_0(y),
\]
and so, a.e.\ one has
\[\frac{|\psi_\varepsilon(x,y)|^2}{\sqrt{x^2+\varepsilon^2 y^2}} \to  \frac{|w(x)u_0(y)|^2}{|x|}.
\]

By Fatou Lemma,
\begin{eqnarray*}
\liminf_{\varepsilon \to  0} \int_{\mathbb R} \int_S 
\frac{|\psi_\varepsilon(x,y)|^2}{\sqrt{x^2+\varepsilon^2 y^2}} \, dydx
& \geq & 
\int_{\mathbb R} \int_S \liminf_{\varepsilon \to  0} 
\frac{|\psi_\varepsilon(x,y)|^2}{\sqrt{x^2+\varepsilon^2 y^2}} \, dydx  \\
& = & 
\int_{\mathbb R} \int_S \frac{|w(x)u_0(y)|^2}{|x|} \, dydx \\ & = & 
\int_{\mathbb R}  \frac{|w(x)|^2}{|x|} \, dydx.
\end{eqnarray*}
Hence, by property \ref{proper1}.\ and the above inequality,
\begin{eqnarray*}
\liminf_{\varepsilon \to  0} b^\varepsilon_\kappa(\psi_\varepsilon) & \geq & 
\liminf_{\varepsilon \to  0} \int_{\mathbb R} \int_S 
\left( |\psi'_\varepsilon + \nabla_\perp \psi_\varepsilon \cdot R y \alpha'|^2  + 
\frac{|\psi_\varepsilon|^2}{\sqrt{x^2+\varepsilon^2 y^2}} \right) \, dydx  \\
& \geq &
\liminf_{\varepsilon \to  0} \int_{\mathbb R} \int_S 
|\psi_\varepsilon' + \nabla_\perp \psi_\varepsilon \cdot R y \alpha'|^2 \, dx dy \\ &+& \liminf_{\varepsilon \to  0}
\int_{\mathbb R} \int_S \frac{|\psi_\varepsilon|^2}{\sqrt{x^2+\varepsilon^2 y^2}}
\, dydx \\
& \geq &
\int_{\mathbb R} \int_S  |w' u_ 0 + w \nabla_\perp u_0 \cdot R y \alpha'|^2 \, dydx + 
\int_{\mathbb R} \int_S  \frac{|wu_0|^2}{|x|} \, dydx\\
& = & 
\int_{\mathbb R} \left(|w'|^2 + (\alpha'(x))^2 C(S) |w|^2 \right) \, dx + \int_{\mathbb R} \frac{|w|^2}{|x|} \, dx.
\end{eqnarray*}
Since it is supposed that
$\liminf_{\varepsilon \to  0}  b^\varepsilon_\kappa(\psi_\varepsilon) < \infty$, it follows that
\[
 \int_{\mathbb R} \frac{|w|^2}{|x|} \, dx < \infty.
 \]
Hence, $w(0^+)=0=w(0^-)$, that is, $w \in \dom b^0_\kappa $, and it also follows that
\[
\liminf_{\varepsilon \to  0} b^\varepsilon_\kappa(\psi_\varepsilon) \geq b^0_\kappa(\psi).
\]

\

\noindent {\bf Step 2.} For $\psi(x, y) = w(x) u_0(y)$, $w \in \dom b^0_\kappa $, take $\psi_\varepsilon = \psi$, for all $ \varepsilon>0$.
So, $\psi_\varepsilon \to  \psi$ in $\LL^2(\mathbb R \times S)$. By monotone convergence, one finds 
\begin{eqnarray*}
\lim_{\varepsilon \to  0}  b^\varepsilon_\kappa(\psi_\varepsilon) & = &
\lim_{\varepsilon \to  0} \int_{\mathbb R} \int_S 
\left( |w' u_ 0 + w \nabla_\perp u_0 \cdot R y \alpha'|^2 + 
\frac{|wu_0|^2}{\sqrt{x^2+\varepsilon^2 y^2}} \right)
\, dydx \\
& = &
\int_{\mathbb R} \int_S \left( |w' u_ 0 + w \nabla_\perp u_0 \cdot R y \alpha'|^2 + 
\frac{|wu_0|^2}{|x|} \right) \, dydx \\
& =& 
\int_{\mathbb R} \left( 
|w'(x)|^2 + \alpha'(x)^2 C(S) |w(x)|^2+ 
\frac{|w(x)|^2}{|x|} \right) \, dx = b^0_\kappa (\psi).
\end{eqnarray*}

Now take $\psi \neq w u_0$, $w \in \dom b^0_\kappa $ and let $\psi_\varepsilon \to  \psi$ in
$\LL^2(\mathbb R \times S)$. 
Then, $\lim_{\varepsilon \to  0}  b^\varepsilon_\kappa(\psi_\varepsilon) = \infty = b^0_\kappa(\psi)$.
In fact, if $\liminf_{\varepsilon \to  0} b^\varepsilon_\kappa(\psi_\varepsilon) < \infty$ one gets, by Step~1, that  $\psi=wu_0$ with $w \in \dom b^0_\kappa $, a contradiction with our hypothesis. 

Summing up, by taking into account Appendix~\ref{appendixGammaSR}, Steps~1 and~2 constitute a proof of the following

\begin{prop2}\label{propGmmConv}
If $\kappa<0$, then $b^\varepsilon_\kappa \sgconv b^0_\kappa$ in $\LL^2(\R\times S)$ as $\varepsilon\to 0$.
\end{prop2}
 
 \

\noindent {\bf Step 3.} Let $\psi\in\LL^2(\R\times S)$ and $\psi_\varepsilon \rightharpoonup \psi$ weakly in $\LL^2(\mathbb R \times S)$.

If $\liminf_{\varepsilon \to  0} b^\varepsilon_\kappa(\psi_\varepsilon) = \infty$, then clearly
$\liminf_{\varepsilon \to  0} b^\varepsilon_\kappa(\psi_\varepsilon) \geq b^0_\kappa(\psi)$.
If 
\[
\liminf_{\varepsilon \to  0} b^\varepsilon_\kappa( \psi_\varepsilon) < \infty,
\]
by taking a subsequence one may suppose that
\[
\liminf_{\varepsilon \to  0} b^\varepsilon_\kappa( \psi_\varepsilon) = 
\lim_{\varepsilon \to  0} b^\varepsilon_\kappa(\psi_\varepsilon) < \infty.
\]
By repeating some of the arguments employed in the steps above, it is found that
$\psi \in{\mathcal H}_0^1(\mathbb R \times S)$,
$\psi'_\varepsilon \rightharpoonup \psi'$ and $\nabla_\perp \psi_\varepsilon \rightharpoonup \nabla_\perp \psi$ weakly in 
$\LL^2(\mathbb R \times S)$. We next divide the argument in three exhaustive cases.

\

{\it Case 1.}
In this case,  suppose that $\psi(x,y)= w(x) u_0(y)$ with $w \in \dom b^0_\kappa $.

Given $a > 0$ define, for $\psi \in{\mathcal H}_0^1(\mathbb R \times S)$,
\begin{eqnarray*}
b^\varepsilon_{\kappa,a} (\psi) &:=& \int_{\mathbb R} \int_S 
\Big(|\psi' + \nabla_\perp \psi \cdot R y \alpha'|^2 \\ &+& \frac{1}{\varepsilon^2} \left( |\nabla_\perp \psi|^2 - \lambda_0 |\psi|^2 \right) +
\frac{|\psi|^2}{a+\sqrt{x^2 + \varepsilon^2 y^2 }}\Big) \, dydx.
\end{eqnarray*}
Since $S$ is a bounded region, it follows that, as $\varepsilon\to0$,
$$\displaystyle\frac{1}{a+\sqrt{x^2 + \varepsilon^2 y^2 }} \to 
\frac{1}{|x|+ a}$$
uniformly in $\mathbb R \times S$, and since $\frac{1}{|x|+a} \in \LL^\infty (\mathbb R \times S)$, one has the weak convergence
$$\frac{1}{|x|+a} \psi_\varepsilon  \rightharpoonup \frac{1}{|x|+a} \psi$$  
in $\LL^2(\mathbb R \times S)$, and so
\begin{eqnarray*}
&\liminf_{\varepsilon \to  0}& b^\varepsilon_{\kappa,a} (\psi_\varepsilon)  \\ & \geq & 
\liminf_{\varepsilon \to  0} \int_{\mathbb R} \int_S 
\left( |\psi'_\varepsilon + \nabla_\perp \psi_\varepsilon \cdot Ry \alpha'|^2 + 
\frac{|\psi_\varepsilon|^2}{a+\sqrt{x^2 + \varepsilon^2 y^2 }} \right) \, dydx \\
& \geq &
\liminf_{\varepsilon \to  0} \int_{\mathbb R} \int_S 
|\psi'_\varepsilon + \nabla_\perp \psi_\varepsilon \cdot R y \alpha'|^2 \, dydx 
\\ &+& \liminf_{\varepsilon \to  0} \int_{\mathbb R} \int_S 
\frac{|\psi_\varepsilon |^2}{|x|+a} \, dydx \\
& \geq & 
\int_{\mathbb R} \int_S 
|\psi' + \nabla_\perp \psi \cdot R y \alpha'|^2 \, dydx + 
\int_{\mathbb R} \int_S \frac{|\psi|^2}{|x|+a} \, dydx \\
& = &
\int_{\mathbb R} \left(|w'(x)|^2  +  (\alpha'(x))^2 C(S) |w(x)|^2 \right) \, dx +
\int_{\mathbb R} \frac{|w(x)|^2}{|x|+a} \, dx.
\end{eqnarray*}

Since 
\[
\frac{|\psi_\varepsilon|^2}{\sqrt{x^2 + \varepsilon^2 y^2 }} \geq
\frac{|\psi_\varepsilon|^2}{a+\sqrt{x^2 + \varepsilon^2 y^2 }},
\]
one has
\begin{eqnarray*}
\liminf_{\varepsilon \to  0} b^\varepsilon_\kappa(\psi_\varepsilon) 
&\geq& 
\liminf_{\varepsilon \to  0} b^\varepsilon_{\kappa,a}(\psi_\varepsilon)
\\ &\geq& 
\int_{\mathbb R} \left(|w'(x)|^2  +  (\alpha'(x))^2 C(S) |w(x)|^2 \right) \, dx +
\int_{\mathbb R} \frac{|w(x)|^2}{|x|+a} \, dx.
\end{eqnarray*}
Since $a > 0$ is arbitrary, 
\begin{eqnarray*}
\liminf_{\varepsilon \to  0} b^\varepsilon_\kappa(\psi_\varepsilon) \geq
\lim_{a \to  0^+} \left(\int_{\mathbb R} \left(|w'(x)|^2  +  
(\alpha'(x))^2 C(S) |w(x)|^2 \right) \, dx +
\int_{\mathbb R} \frac{|w(x)|^2}{|x|+a} \, dx \right).
\end{eqnarray*}
By monotone convergence
\begin{eqnarray*}
\liminf_{\varepsilon \to  0} &b^\varepsilon_\kappa(\psi_\varepsilon) & \\ &\geq&
\int_{\mathbb R} \left(|w'(x)|^2  +  (\alpha'(x))^2 C(S) |w(x)|^2 \right) \, dx +
\int_{\mathbb R} \frac{|w(x)|^2}{|x|} \, dx \\ &=& b^0_\kappa(\psi) = b^0_\kappa (w).
\end{eqnarray*}

\

{\it Case 2.}
In this case assume that $\psi \notin A:= \{w u_0: w \in{\mathcal H}^1(\mathbb R)\}$. By definition,
$b^0_\kappa(\psi)=\infty$.
Let $P$ denote the orthogonal projection onto $A^\perp$ and let $[u_0]$ be the subspace generated by $u_0$. Then, 
$\|P\psi\| > 0$ and $P \psi_\varepsilon \rightharpoonup P\psi$, and so
\[
\liminf_{\varepsilon \to  0} \|P\psi_\varepsilon\| \geq \|P\psi\| > 0.
\]
Observe that 
\[
\liminf_{\varepsilon \to  0} b^\varepsilon_\kappa (\psi_\varepsilon) \geq 
\liminf_{\varepsilon \to  0}
\frac{1}{\varepsilon^2} \int_{\mathbb R} \int_S \left(|\nabla_\perp \psi_\varepsilon|^2 - \lambda_0 |\psi_\varepsilon|^2\right) dydx.
\]

Let us estimate the limit on the right hand side.
For $\phi \in{\mathcal H}^1(\mathbb R)$, denote by $\phi^{(0)}$  the component of $\phi$ in the subspace $[u_0]$.
Let $Q$ be the orthogonal projection onto $[u_0]^\bot$ in ${\mathcal H}_0^1(S)$.

Let that $\lambda_1>\lambda_0$ is the second eigenvalue of $-\Delta_\perp$ in~$S$. Note that for $\varphi \in [u_0]^{\bot}$ one has
$\int_S | \nabla \varphi|^2 \geq \lambda_1 \int_S |\varphi|^2$, and so
\begin{eqnarray*}
&   
\displaystyle{\int_{\mathbb R}}& \int_S 
\left( |\nabla_\perp \psi_\varepsilon|^2 - \lambda_0 |\psi_\varepsilon|^2 \right)\, dydx \\
& = & 
  \int_\mathbb R
\left( \| \nabla_\perp \psi_\varepsilon(x) \|^2_{L^2(S)} - \lambda_0 \|\psi_\varepsilon(x)\|^2_{L^2(S)} \right)\, dx \\
& = &
  \int_\mathbb R
\left( \|\psi_\varepsilon(x)\|^2_{{\mathcal H}_0^1(S)} -(\lambda_0+1) \|\psi_\varepsilon(x)\|^2_{L^2(S)}\right) \, dx \\
& = &
  \int_\mathbb R
\left( \| Q \psi_\varepsilon(x)\|^2_{{\mathcal H}_0^1(S)}  +  \| \psi_\varepsilon^{(0)}\|^2_{{\mathcal H}_0^1(S)} 
-(\lambda_0+1) \| \psi_\varepsilon(x)\|^2_{L^2(S)}\right) \, dx \\
& = &
  \int_{\mathbb R} \left(
\| \nabla_ y Q \psi_\varepsilon(x)\|^2_{L^2(S)} + \|Q \psi_\varepsilon(x)\|^2_{L^2(S)} + 
\|\nabla_\perp  \psi_\varepsilon^{(0)}(x)\|^2_{L^2(S)} \right. \\
& + & \left.
\|\psi_\varepsilon^{(0)}(x)\|^2_{L^2(S)} -\lambda_0
\| \psi_\varepsilon(x)\|^2_{L^2(S)} - \| \psi_\varepsilon(x)\|^2_{L^2(S)} \right) \, dx \\
& = &
  \int_\mathbb R
\left( \| \nabla_\perp Q \psi_\varepsilon(x)\|^2_{L^2(S)} + 
\| \nabla_\perp \psi_\varepsilon^{(0)}(x)\|^2_{L^2(S)} - \lambda_0
\| \psi_\varepsilon(x)\|^2_{L^2(S)}\right) \, dx \\
& \geq &
  \int_\mathbb R
\left( \lambda_ 1 \| Q \psi_\varepsilon(x)\|^2_{L^2(S)} + \lambda_0 
\| \psi_\varepsilon^{(0)}(x)\|^2_{L^2(S)}
- \lambda_0 \| \psi_\varepsilon(x)\|^2_{L^2(S)} \right) \, dx \\
& = &
  \int_\mathbb R
(\lambda_1 - \lambda_0) \| Q \psi_\varepsilon(x) \|^2_{L^2(S)}\, dx \\
& = &
{(\lambda_1-\lambda_0)} \|P\psi_\varepsilon\|^2 \geq
{(\lambda_1-\lambda_0)} \|P\psi\|^2.
\end{eqnarray*}

By recalling that $\lambda_0 < \lambda_1$ and $\|P\psi\| > 0$, one obtains
\[
\liminf_{\varepsilon \to  0} \displaystyle\frac{1}{\varepsilon^2} 
\int_{\mathbb R} \int_S 
\left(|\nabla_\perp \psi_\varepsilon|^2 - \lambda_0 |\psi_\varepsilon|^2\right) \, dydx \geq
\lim_{\varepsilon \to  0} \frac{(\lambda_1-\lambda_0)}{\varepsilon^2} \|P\psi\|^2 = \infty.
\]
Hence,
\[
\liminf_{\varepsilon \to  0} b^\varepsilon_\kappa (\psi_\varepsilon) \geq b^0_\kappa(\psi).
\]

\

{\it Case 3.} 
In this case assume that $v = w u_0$ with $w \in{\mathcal H}^1(\mathbb R) \backslash \dom b^0_\kappa $.
Then, $\liminf_{\varepsilon \to  0} b^\varepsilon_\kappa(\psi_\varepsilon) = \infty.$
In fact, if $\liminf_{\varepsilon \to  0} b^\varepsilon_\kappa(\psi_\varepsilon) < \infty$,
by passing to a subsequence, as in {\it Case 1},
\[
\infty > \lim_{\varepsilon \to  0} b^\varepsilon_\kappa(\psi_\varepsilon) \geq 
\lim_{a \to  0} \left( \int_{\mathbb R} |w' + \nabla_\perp w \cdot R y \alpha'|^2 \, dx +
\int_{\mathbb R}  \frac{|w|^2}{|x|+a} \, dx \right),
\]
which is impossible, since $w \in{\mathcal H}^1(\mathbb R) \backslash \dom b^0_\kappa $ and so
\[
\lim_{a \to  0} \int_{\mathbb R}  \frac{|w|^2}{|x|+a}  \, dx  = \infty.
\]
This finishes Step~3.

\

Finally, combine Step~3, Proposition~\ref{propGmmConv} and statements (j) and~(jjj) in Appendix~\ref{appendixGammaSR} to conclude the proof of Theorem~\ref{teorRepulsive}.

\appendix 
\numberwithin{equation}{section}

\section{The one-dimensional H-atom}\label{1dHatom}
In this appendix we  review a characterization of the self-adjoint extensions related to the {\sc1D} hydrogen atom, as well as some of their properties we make use in this paper  \cite{deOV}. The initial {\sc1D} hermitian operator is $\dot H$ defined by equation \eqref{defHponto}, its deficiency indices are equal to~2, and its adjoint operator $\dot H^*$ has the same action as $\dot H$ but domain 
\[\dom \dot H^*  =  
\left\{ \phi \in \LL^2(\R\setminus\{0\})  : \phi, \phi' \in \mathrm{AC} (\R\setminus \{0\} ),
\left(-   \phi'' - \frac{\kappa}{|x|} \phi\right)  \in \LL^2(\R) \right\}.
\]

If $\phi \in \dom  \dot H^*$, then it turns out \cite{deOV} that the lateral limits $\phi(0^\pm):=
\lim_{x\to 0^\pm}\phi(x)$ exist (and are finite) and, for $\kappa\ne0$,
\begin{equation}\label{phiTil} 
  \tilde{\phi} (0^{\pm}) :=   \lim_{x \rightarrow 0^{\pm}} 
\left( \phi'(x) \pm \kappa \phi(x) \ln( \pm |\kappa| x) \right)
\end{equation}
exist and are finite as well.  The $2 \times 2$ unitary matrices $\hat U$ characterize the self-adjoint extensions $ H_{\hat \eta}$ of $\dot H$, so that for each of such matrices it is found that $\dom  H_{\hat \eta}$ is composed of  
$\psi \in \dom\dot H^*$
so that \[
(I - \hat \eta) \left(\begin{array}{cc}
\tilde{\psi}(0^+) \\
\tilde{\psi}(0^-)
          \end{array}\right) =
- i (I + \hat \eta) \left(\begin{array}{cc}
-\psi(0^+) \\
\psi(0^-)
          \end{array}\right),
\] where $I$ denotes the $2\times 2$ identity matrix, and we have explicitly got the boundary conditions characterizing the  self-adjoint extensions of $\dot H$, that is, the operator candidates for the description of the one-dimensional hydrogen atom. For instance, the choice $\hat U=I$ leads to the operator 
\begin{equation}\label{DirichletHamiltonian}
(H_D\phi)(x) = - \phi''(x) - \frac{\kappa}{|x|} \phi(x)
\end{equation} with Dirichlet boundary condition $\phi(0^+)=0=\phi(0^-)$.

\section{The operator $H_\kappa^0$}\label{appHkappa0}

We show that the self-adjoint operator $H_\kappa^0$ in $\R$, defined in Section~\ref{sectSetup}, 
corresponds to the self-adjoint extension $H_D$ (see \eqref{DirichletHamiltonian}) of $\dot{H}$. To simplify expressions, without loss we assume that $\alpha'(x)=0$.
Consider the self-adjoint operator $(\mathbf h w)(x) = - w''(x)$, 
$$\dom {\mathbf h} = \left\{ w \in \hil^2(\mathbb R): w(0^+) = 0 = w(0^-)\right\}.$$

This operator is associated with the sesquilinear form 
$$b_{\mathbf h}(w, \varphi)= \langle w', \varphi' \rangle, \hspace{1cm} w, \varphi \in \dom b_{\mathbf h},$$
with $\dom b_{\mathbf h} = \mathcal D_0= \left\{w  \in \hil^1(\mathbb R): w(0^+) = 0 = w(0^-)\right\},$
that is,
$\dom b_{\mathbf h}$ is the  form domain of ${\mathbf h}$ (see also \cite{ISTQD}).

The form domain of the multiplication operator $V(x) = \displaystyle - \frac{\kappa}{|x|}$,
$x,\kappa \in \mathbb R$, $\kappa \neq 0$,
is
$$\dom b_V = \left\{ w \in \LL^2(\mathbb R): \frac{w(x)}{\sqrt{x}} \in \LL^2(\mathbb R) \right\}.$$

\

\underline{Claim}:  $V$ is $b_{\mathbf h}$-bounded and the  $b_{\mathbf h}$-bound of $V$ is less than $1$.
\begin{proof}
Initially, note that $\dom b_{\mathbf h} \subset \dom b_V$. Given $a> 0$, choose $\delta > 0$ so that
$\displaystyle \frac{1}{|x|} \leq \frac{a}{4} \frac{1}{|x|^2}$ for all $|x| \leq \delta$.
By Hardy's Inequality, for $w\in \dom b_{\mathbf h}$ one has
\begin{eqnarray*}
|b_V (w) | & = & |\kappa| \int_\mathbb R \frac{|w(x)|^2}{|x|} dx \\
& = &
  |\kappa| \int_{|x| \leq \delta} \frac{|w(x)|^2}{|x|} dx +   |\kappa| \int_{|x| \geq \delta} \frac{|w(x)|^2}{|x|} dx \\
& \leq &
 |\kappa| \frac{a}{4} \int_\mathbb R \frac{|w(x)|^2}{|x|^2} dx + \frac{ |\kappa|}{\delta} \int_\mathbb R |w(x)|^2 dx \\
& \leq &
|\kappa| a \int_\mathbb R |w(x)'|^2 dx +   \frac{|\kappa|}{\delta} \int_\mathbb R |w(x)|^2 dx \\
& = &
  a\,|\kappa|\, b_{\mathbf h}(w) +  \frac{|\kappa|}{\delta} \| w\|^2.
\end{eqnarray*}
Since $a>0$ was arbitrary, the result follows. 
\end{proof}

Since
\[
b^0_\kappa(w) = b_{\mathbf h}(\varphi, w) + b_V( w) 
\]with $\dom b^0_\kappa =\dom  b_{\mathbf h}$, by KLMN Theorem $\cite{ISTQD}$, there exists a unique self-adjoint operator
$H$ with $\dom H  \subset  \dom b_{\mathbf h}$, whose form domain is $\dom b_{\mathbf h}$,  such that
\begin{equation}\label{aaaa} 
\langle \varphi , H w \rangle = b^0_\kappa(\varphi,w),\qquad
\forall \varphi \in \dom b_{\mathbf h}, \forall w \in \dom H.
\end{equation}
It is also known that $\dom H$ is dense in $\dom b_{\mathbf h}$ with respect to the inner product
$\langle \cdot, \cdot \rangle_+ := b^0_\kappa (\cdot, \cdot)  +c \langle \cdot, \cdot \rangle$, with  $\langle \cdot, \cdot \rangle$ denoting the usual inner product  in $\LL^2(\mathbb R)$.

Note that $\dom H_D$ is  dense in $(\dom b_{\mathbf h}, \langle \cdot , \cdot \rangle_+)$.
In fact, since $\dom b_{\mathbf h}$ is the form domain of ${\mathbf h}$,
for each $w \in \dom b_{\mathbf h}$ there is a sequence  
$(w_n) \subset \dom {\mathbf h} = \dom H_D$
so that
\begin{equation}\label{bbbb} 
b_{\mathbf h}(w_n - w) + \|w_n - w \|^2 \rightarrow  0.
\end{equation}

Now we pick $a> 0$ as above, and so, by Hardy's Inequality,
\begin{equation}\label{cccc} 
|b_V (w) | \leq d \left( b_{\bf h}(w) + || w||^2 \right),
\end{equation}
where
$d = \max\{a|\kappa|, |\kappa|/\delta\}$.
Thus, by~\eqref{bbbb} and~\eqref{cccc}, $b_V(w_n - w)  \rightarrow 0$. Hence
$$\langle w_n-w, w_n-w \rangle_+ \rightarrow 0,$$
i.e., $\dom H_D$ is dense in $(\dom b_{\mathbf h}, \langle \cdot , \cdot \rangle_+)$.
An integration by parts
shows that condition~\eqref{aaaa} also holds with $H$ replaced by $H_D$; so, by uniqueness, 
it follows that $H = H_D$, that is, $H_D$ is the operator associated with the form~$b^0_\kappa$.

\section{$\Gamma$ and strong resolvent convergences}\label{appendixGammaSR}
In this appendix we recall, in an appropriate and practical way, the definition of $\Gamma$-convergence of quadratic forms in Hilbert spaces and its relation to the strong resolvent convergence of the subsequent self-adjoint operators. A more detailed review can be found in \cite{GammaConvT}, whereas a full account appears in~\cite{DalMaso}. 

Let $\hil$ be a Hilbert space and $\overline \R=\R\cup\{+\infty\}$. The sequence of functionals $f_\varepsilon:\hil\to \overline{\R}$ strongly $\Gamma$-converges to $f$ as $\varepsilon\to0$ (denoted by $f_\varepsilon\sgconv f$) if the following two conditions hold:
\begin{itemize}
\item[(i)] For every $\zeta\in\hil$ and every $\zeta_\varepsilon\to\zeta$ in $\hil$ one has
\begin{equation}\label{gammaConv1} f(\zeta)\le \liminf_{\varepsilon\to0} f_\varepsilon(\zeta_\varepsilon).
\end{equation}
\item[(ii)] For each $\zeta\in\hil$, there exists a sequence $\zeta_\varepsilon\to\zeta$ in
$\hil$ such that
\begin{equation}\label{gammaConv2} f(\zeta)=\lim_{\varepsilon\to0} f_\varepsilon(\zeta_\varepsilon).
\end{equation}
\end{itemize}

 If the strong con\-ver\-gence $\zeta_\varepsilon\to\zeta$ is replaced by weak con\-ver\-gence $\zeta_\varepsilon\wseta\zeta$ in (i) and (ii) above, then one says that $f_\varepsilon$ weakly $\Gamma$-converges to $f$, which is denoted by $f_\varepsilon\wgconv f$.

 Let $b^\varepsilon,b$ be positive (or, more generally, uniformly lower bounded)
closed sesqui\-lin\-e\-ar forms in the Hilbert space $\hil$ taking values in $\overline \R$, and
$T^\varepsilon,T$ the corresponding associated positive self-adjoint operators. Then the following
statements are equivalent:
\begin{itemize}
\item[(j)] $b^\varepsilon\sgconv b$ and, for each $\zeta\in\hil$, $b(\zeta)\le
\liminf_{\varepsilon\to0} b^\varepsilon(\zeta_\varepsilon)$, for all $\zeta_\varepsilon\wseta
\zeta$ in~$\hil$.
\item[(jj)] $b^\varepsilon+\lambda \sgconv b+\lambda$ and $b^\varepsilon+\lambda \wgconv b+\lambda$,
for some
$\lambda\ge0$ (and so for all $\lambda\ge0$).
\item[(jjj)] $T^\varepsilon$ converges to $T$ in the strong resolvent sense in
$\hil_0=\overline{\dom T}\subset\hil$, that is, for all $\lambda\in \C\setminus(-\infty,0]$,
\begin{equation}\label{resConvTeps} 
\lim_{\varepsilon\to0} R_{-\lambda}(T^\varepsilon)\zeta = R_{-\lambda}(T)P_0\zeta,\quad
\forall\zeta\in\hil,
\end{equation} where $P_0$ is the orthogonal projection onto~$\hil_0$.
\end{itemize}

\

 \subsubsection*{Acknowledgments} {\small We thank an anonymous referee for careful reading our manuscript and indicating  important corrections. CRdO acknowledges partial
support from CNPq (Brazil). AAV was  supported by CAPES (Brazil).}

\

\end{document}